\documentclass{article}
\usepackage{amssymb}
\usepackage{amsmath}
\usepackage{cite}

\input{tcilatex}
\begin{document}

\title{Hagen-Hurley equations and the $W$ boson}
\author{Andrzej Okni\'{n}ski \\
Politechnika \'{S}wi\c{e}tokrzyska, \\
Al. 1000-lecia PP 7, 25-314 Kielce, Poland}
\maketitle

\begin{abstract}
We proceed with our study of the Hagen-Hurley equations describing spin one
bosons. In this work, we describe a general decay of the Hagen-Hurley boson.
It is important that the transformation conserves spin $s=1$ of the decaying
boson and provides information about the kinematics of the decay. We explain
the high instability of the Hagen-Hurley particle and  identify it as the $W$
boson. 
\end{abstract}

\section{Introduction}
\label{introduction}

The intermediate vector bosons, $W^{\pm }$ as well as $Z^{0}$, are extremely
short-lived particles with a half-life of about $3\times 10^{-25}$ s \cite%
{Workman2022}. Very recently, an analysis of high-precision measurement of the $%
W$ boson mass suggests that the mass is slightly higher than the Standard
Model predicts \cite{CDF2022}. This result hints at the need for reconsideration
of the Standard Model.

This work aims to propose a relativistic equation that could
effectively describe the $W$ boson. 

We base our approach on spin one Hagen-Hurley equation \cite%
{HagenHurley1970,Hurley1971,Hurley1974,Beckers1995a,Beckers1995b}. 
In Section \ref{rearrangement}, for the sake of completeness, 
we rewrite the Hagen-Hurley equations as two coupled 
Dirac equations involving higher-order spinors as in \cite{Okninski2015}. 

Then, in Section \ref{decay}, we
describe decays of \textit{real} $W$ bosons, such as appearing in the top quark
decay or produced in collisions of protons and antiprotons. 
More precisely, we transform two coupled Dirac equations involving
non-standard spinors, obtained in Section \ref{rearrangement}, into two
Dirac equations for two fermions, generalizing our earlier work \cite%
{Okninski2016}. It is important that this transformation conserves spin $s=1$
of the decaying boson and provides information about the kinematics of the
decay, as described in Section \ref{kinematics}.

In what follows, we are using definitions and conventions of Ref. \cite
{Okninski2015}.

\section{Rearrangement of the Hagen-Hurley equations as two coupled Dirac equations}
\label{rearrangement}

We start with the spinor formulation of the Hagen-Hurley equations, cf. \cite%
{Dirac1936} or Subsection 6 ii) in \cite{Lopuszanski1978} or Eqs. (18), (19)
in \cite{Okninski2015}, see also Eqs. (2.1), (2.24) in \cite{Beckers1995a}
for the $7\times 7$\ form. These equations violate parity $P$ \cite%
{Lopuszanski1978}, where $P:x^{0}\rightarrow x^{0},\ x^{i}\rightarrow
-x^{i}\ \left( i=1,2,3\right) $. Since parity is violated in weak
interactions \cite{Thomson2013,Donoghue2014}, these equations can describe
weakly interacting particles.

We write one of these equations (Eq. (19) of Ref. \cite{Okninski2015}), in the 
form: 
\begin{subequations}
\label{HH}
\begin{gather}
\left. 
\begin{array}{l}
p_{\ \ \dot{B}}^{A}\ \zeta _{A\dot{D}}=m\chi _{\dot{B}\dot{D}}\smallskip \\ 
p_{A}^{\ \ \dot{D}}\ \chi _{\dot{B}\dot{D}}=-m\zeta _{A\dot{B}}%
\end{array}%
\right\}  \label{HH1} \\
\chi _{\dot{B}\dot{D}}=\chi _{\dot{D}\dot{B}}  \label{s=1}
\end{gather}%
where Eq. (\ref{s=1}) is the spin-$1$ constraint \cite{Lopuszanski1978}. In
Eqs. (\ref{HH1}) we have $p^{A\dot{B}}=\left( \sigma ^{0}p^{0}+%
\overrightarrow{\sigma }\cdot \overrightarrow{p}\right) ^{A\dot{B}}$, $%
p^{\mu }=i\frac{\partial }{\partial x_{\mu }}$, where $\sigma ^{k}$ $\left(
k=1,2,3\right) $ are the Pauli matrices, and $\sigma ^{0}$ is the $2\times 2$
unit matrix. Equations (\ref{HH}), which were first proposed by Dirac \cite%
{Dirac1936}, can be written in the $7\times 7$\ Hagen-Hurley form: 
\end{subequations}
\begin{equation}
\beta ^{\mu }p_{\mu }\Psi =m\Psi ,  \label{HHform}
\end{equation}%
the $\beta ^{\mu }$ matrices given in \cite{Beckers1995a}, with $\Psi
=\left( \chi _{\dot{1}\dot{1}},\chi ,\chi _{\dot{2}\dot{2}},\zeta _{1\dot{1}%
},\zeta _{1\dot{2}},\zeta _{2\dot{1}},\zeta _{2\dot{2}}\right) ^{T}$, where $%
\chi \overset{df}{=}\chi _{\dot{1}\dot{2}}=\chi _{\dot{2}\dot{1}}$ and $^{T}$
stands for transposition of a matrix.

Equations (\ref{HH1}) in explicit form read: 
\begin{subequations}
\label{HH2}
\begin{eqnarray}
&&\left. 
\begin{array}{rcr}
-\left( p^{1}+ip^{2}\right) \chi _{\dot{1}\dot{1}}-\left( p^{0}-p^{3}\right)
\chi _{\dot{2}\dot{1}} & = & -m\zeta _{1\dot{1}} \\ 
\left( p^{0}+p^{3}\right) \chi _{\dot{1}\dot{1}}+\left( p^{1}-ip^{2}\right)
\chi _{\dot{2}\dot{1}} & = & -m\zeta _{2\dot{1}} \\ 
-\left( p^{1}-ip^{2}\right) \zeta _{1\dot{1}}-\left( p^{0}-p^{3}\right)
\zeta _{2\dot{1}} & = & m\chi _{\dot{1}\dot{1}} \\ 
\left( p^{0}+p^{3}\right) \zeta _{1\dot{1}}+\left( p^{1}+ip^{2}\right) \zeta
_{2\dot{1}} & = & m\chi _{\dot{2}\dot{1}}%
\end{array}%
\right\}  \label{HH2a} \\
&&\left. 
\begin{array}{rcr}
-\left( p^{1}+ip^{2}\right) \chi _{\dot{1}\dot{2}}-\left( p^{0}-p^{3}\right)
\chi _{\dot{2}\dot{2}} & = & -m\zeta _{1\dot{2}} \\ 
\left( p^{0}+p^{3}\right) \chi _{\dot{1}\dot{2}}+\left( p^{1}-ip^{2}\right)
\chi _{\dot{2}\dot{2}} & = & -m\zeta _{2\dot{2}} \\ 
-\left( p^{1}-ip^{2}\right) \zeta _{1\dot{2}}-\left( p^{0}-p^{3}\right)
\zeta _{2\dot{2}} & = & m\chi _{\dot{1}\dot{2}} \\ 
\left( p^{0}+p^{3}\right) \zeta _{1\dot{2}}+\left( p^{1}+ip^{2}\right) \zeta
_{2\dot{2}} & = & m\chi _{\dot{2}\dot{2}}%
\end{array}%
\right\}  \label{HH2b}
\end{eqnarray}
\end{subequations}
These equations are coupled due to the condition $\chi _{\dot{1}\dot{2}%
}=\chi _{\dot{2}\dot{1}}$ which ensures that $s=1$.

\section{Decays of real spin-$1$ Hagen-Hurley bosons}
\label{decay}

We realize that solutions of two Dirac equations (\ref{HH2}) are
non-standard since they involve higher-order spinors rather than spinors $%
\xi _{A}$,$\ \eta _{\dot{B}}$. There are several possibilities to reduce
these higher-order spinors via the de Broglie method of fusion \cite%
{deBroglie1943,Corson1953}.

Attempting to describe leptonic decays of the $W$ boson, $W^{-}\rightarrow l+%
\bar{\nu}_{l}$, we made the following substitution \cite{Okninski2016}: 
\begin{subequations}
\label{SUB1}
\begin{eqnarray}
\chi _{\dot{B}\dot{D}}\left( x\right) &=&\eta _{\dot{B}}\left( x\right)
\alpha _{\dot{D}}\left( x\right) ,  \label{sub1a} \\
\zeta _{A\dot{B}}\left( x\right) &=&\xi _{A}\left( x\right) \alpha _{\dot{B}%
}\left( x\right) ,  \label{sub1b}
\end{eqnarray}
\end{subequations}
where $\alpha _{\dot{A}}\left( x\right) $ is the Weyl spinor, describing
massless neutrinos, while $\eta _{\dot{B}}\left( x\right) $, and $\xi _{A}\left(
x\right) $ are the Dirac spinors. Since $\chi _{\dot{B}\dot{D}}\left(
x\right) \neq \chi _{\dot{D}\dot{B}}\left( x\right) $ we had to assume that
the spin of the decaying boson belonged to $0\oplus 1$ space. We suggested that
our formalism can describe a decay of a \textit{virtual} $W^{-}$ boson into
a lepton and antineutrino in the mixed beta decay \cite{Okninski2016}.

On the other hand, to describe the decay of a \textit{rea}l $W$ boson, we carry
out another, conserving spin, substitution:
\begin{subequations}
\label{SUB2}
\begin{eqnarray}
\chi _{\dot{B}\dot{D}}\left( x\right) &=&\eta _{\dot{B}}\left( x\right)
\alpha _{\dot{D}}\left( x\right) +\alpha _{\dot{B}}\left( x\right) \eta _{%
\dot{D}}\left( x\right) ,  \label{sub1c} \\
\zeta _{A\dot{B}}\left( x\right) &=&\xi _{A}\left( x\right) \alpha _{\dot{B}%
}\left( x\right) +\lambda _{A}\left( x\right) \eta _{\dot{B}}\left( x\right)
.  \label{sub2c}
\end{eqnarray}
\end{subequations}
Note that in this case $\chi _{\dot{1}\dot{2}}=\chi _{\dot{2}\dot{1}}$ and
thus $s=1$. Both terms in $\chi _{\dot{B}\dot{D}}\left( x\right) $\ are
necessary to ensure $s=1$\ condition.

Substituting (\ref{SUB2}) into Eqs. (\ref{HH2}) and rearranging terms we
obtain:
\begin{eqnarray}
&&\left. 
\begin{array}{rcr}
-\left( p^{1}+ip^{2}\right) \eta _{\dot{1}}\alpha _{\dot{A}}-\left(
p^{0}-p^{3}\right) \eta _{\dot{2}}\alpha _{\dot{A}} & = & -m\xi _{1}\alpha _{%
\dot{A}} \\ 
\left( p^{0}+p^{3}\right) \eta _{\dot{1}}\alpha _{\dot{A}}+\left(
p^{1}-ip^{2}\right) \eta _{\dot{2}}\alpha _{\dot{A}} & = & -m\xi _{2}\alpha
_{\dot{A}} \\ 
-\left( p^{1}-ip^{2}\right) \xi _{1}\alpha _{\dot{A}}-\left(
p^{0}-p^{3}\right) \xi _{2}\alpha _{\dot{A}} & = & m\eta _{\dot{1}}\alpha _{%
\dot{A}} \\ 
\left( p^{0}+p^{3}\right) \xi _{1}\alpha _{\dot{A}}+\left(
p^{1}+ip^{2}\right) \xi _{2}\alpha _{\dot{A}} & = & m\eta _{\dot{2}}\alpha _{%
\dot{A}}%
\end{array}%
\right\} ,  \label{1a} \\
&&\left. 
\begin{array}{rcr}
-\left( p^{1}+ip^{2}\right) \alpha _{\dot{1}}\eta _{\dot{B}}-\left(
p^{0}-p^{3}\right) \alpha _{\dot{2}}\eta _{\dot{B}} & = & -m\lambda _{1}\eta
_{\dot{B}} \\ 
\left( p^{0}+p^{3}\right) \alpha _{\dot{1}}\eta _{\dot{B}}+\left(
p^{1}-ip^{2}\right) \alpha _{\dot{2}}\eta _{\dot{B}} & = & -m\lambda
_{2}\eta _{\dot{B}} \\ 
-\left( p^{1}-ip^{2}\right) \lambda _{1}\eta _{\dot{B}}-\left(
p^{0}-p^{3}\right) \lambda _{2}\eta _{\dot{B}} & = & m\alpha _{\dot{1}}\eta
_{\dot{B}} \\ 
\left( p^{0}+p^{3}\right) \lambda _{1}\eta _{\dot{B}}+\left(
p^{1}+ip^{2}\right) \lambda _{2}\eta _{\dot{B}} & = & m\alpha _{\dot{2}}\eta
_{\dot{B}}%
\end{array}%
\right\} ,  \label{1b}
\end{eqnarray}%
where $\dot{A},\ \dot{B}=\dot{1},\ \dot{2}$.

Assuming solutions as:%
\begin{eqnarray}
\alpha _{\dot{B}}\left( x\right) &=&\hat{\alpha}_{\dot{B}}e^{-ik\cdot x},%
\text{ }\lambda _{A}\left( x\right) =\hat{\lambda}_{A}e^{-ik\cdot x},\
k^{\mu }k_{\mu }=m_{1}^{2},  \label{sol1} \\
\eta _{\dot{B}}\left( x\right) &=&\hat{\eta}_{\dot{B}}e^{-iq\cdot x},\text{ }%
\xi _{A}\left( x\right) =\hat{\xi}_{A}e^{-iq\cdot x},\ q^{\mu }q_{\mu
}=m_{2}^{2},  \label{sol2}
\end{eqnarray}%
we get two Dirac equations for spinors $\left( \xi _{A},\ \eta _{\dot{B}%
}\right) ^{T}$ and $\left( \lambda _{C},\ \alpha _{\dot{D}}\right) ^{T}$:

\begin{eqnarray}
&&\left. 
\begin{array}{rcr}
-\left( \tilde{p}^{1}+i\tilde{p}^{2}\right) \eta _{\dot{1}}\left( x\right)
-\left( \tilde{p}^{0}-\tilde{p}^{3}\right) \eta _{\dot{2}}\left( x\right) & =
& -m\xi _{1}\left( x\right) \\ 
\left( \tilde{p}^{0}+\tilde{p}^{3}\right) \eta _{\dot{1}}\left( x\right)
+\left( \tilde{p}^{1}-i\tilde{p}^{2}\right) \eta _{\dot{2}}\left( x\right) & 
= & -m\xi _{2}\left( x\right) \\ 
-\left( \tilde{p}^{1}-i\tilde{p}^{2}\right) \xi _{1}\left( x\right) -\left( 
\tilde{p}^{0}-\tilde{p}^{3}\right) \xi _{2}\left( x\right) & = & m\eta _{%
\dot{1}}\left( x\right) \\ 
\left( \tilde{p}^{0}+\tilde{p}^{3}\right) \xi _{1}\left( x\right) +\left( 
\tilde{p}^{1}+i\tilde{p}^{2}\right) \xi _{2}\left( x\right) & = & m\eta _{%
\dot{2}}\left( x\right)%
\end{array}%
\right\}  \label{Dirac1a} \\
&&\left. 
\begin{array}{rcr}
-\left( \bar{p}^{1}+i\bar{p}^{2}\right) \alpha _{\dot{1}}\left( x\right)
-\left( \bar{p}^{0}-\bar{p}^{3}\right) \alpha _{\dot{2}}\left( x\right) & =
& -m\lambda _{1}\left( x\right) \\ 
\left( \bar{p}^{0}+\bar{p}^{3}\right) \alpha _{\dot{1}}\left( x\right)
+\left( \bar{p}^{1}-i\bar{p}^{2}\right) \alpha _{\dot{2}}\left( x\right) & =
& -m\lambda _{2}\left( x\right) \\ 
-\left( \bar{p}^{1}-i\bar{p}^{2}\right) \lambda _{1}\left( x\right) -\left( 
\bar{p}^{0}-\bar{p}^{3}\right) \lambda _{2}\left( x\right) & = & m\alpha _{%
\dot{1}}\left( x\right) \\ 
\left( \bar{p}^{0}+\bar{p}^{3}\right) \lambda _{1}\left( x\right) +\left( 
\bar{p}^{1}+i\bar{p}^{2}\right) \lambda _{2}\left( x\right) & = & m\alpha _{%
\dot{2}}\left( x\right)%
\end{array}%
\right\}  \label{Dirac1b}
\end{eqnarray}%
with rescaled momentum operators:

\begin{equation}
\tilde{p}^{\mu }=i\tfrac{\partial }{\partial x_{\mu }}+k^{\mu },\ \bar{p}%
^{\mu }=i\tfrac{\partial }{\partial x_{\mu }}+q^{\mu }.  \label{definitions}
\end{equation}

\section{Kinematics of decay of the real spin-$1$ Hagen-Hurley bosons}
\label{kinematics}

The forms (\ref{sol1}), and (\ref{sol2}) solve equations (\ref{Dirac1a}), and (\ref%
{Dirac1b}) provided that:%
\begin{equation}
\left( k^{\mu }+q^{\mu }\right) \left( k_{\mu }+q_{\mu }\right) =m^{2}.
\label{condition}
\end{equation}%
We thus define a new four-vector:
\begin{equation}
p^{\mu }\overset{df}{=}k^{\mu }+q^{\mu },  \label{new}
\end{equation}%
and we have $p^{\mu }p_{\mu }=m^{2}$ where $m$ is the decaying boson mass.

Taking into account Eqs. (\ref{sol1}), (\ref{sol2}), and (\ref{condition}) we
obtain two-body decay kinematic equality:
\begin{equation}
\left. 
\begin{array}{l}
m_{1}^{2}+m_{2}^{2}+2\left( k^{0}q^{0}-\overrightarrow{k}\cdot 
\overrightarrow{q}\right) =m^{2}, \\ 
k^{0}=+\sqrt{\left\vert \overrightarrow{k}\right\vert ^{2}+m_{1}^{2}},\
q^{0}=+\sqrt{\left\vert \overrightarrow{q}\right\vert ^{2}+m_{2}^{2}}.
\end{array}
\right\}   \label{equality-1a}
\end{equation}
It follows that transition from Eqs. (\ref{HH2}) to (\ref{Dirac1a}), (\ref
{Dirac1b}), via the substitution (\ref{SUB2}), can be interpreted as the decay
of a real (not virtual) spin one boson with mass $m$ into two spin one-half
fermions with four-momenta $k^{\mu }$, $q^{\mu }$, and masses $m_{1}$, $m_{2}$
because the inequality: 
\begin{subequations}
\label{INEQ}
\begin{equation}
m^{2}-m_{1}^{2}-m_{2}^{2}=2\left( k^{0}q^{0}-\overrightarrow{k}\cdot 
\overrightarrow{q}\right) >0,  \label{inequality-1}
\end{equation}
is fulfilled.

Indeed, we have 
\begin{equation}
k^{0}q^{0}-\overrightarrow{k}\cdot \overrightarrow{q}=\sqrt{\left\vert 
\overrightarrow{k}\right\vert ^{2}+m_{1}^{2}}\;\sqrt{\left\vert 
\overrightarrow{q}\right\vert ^{2}+m_{2}^{2}}-\left\vert \overrightarrow{k}%
\right\vert \left\vert \overrightarrow{q}\right\vert \cos \varphi >0,
\label{inequality-2}
\end{equation}
\end{subequations}
where we have used the definition of the scalar product $\overrightarrow{k}\cdot 
\overrightarrow{q}$ and $\varphi $ is an angle between vectors $
\overrightarrow{k}$ and $\overrightarrow{q}$.

Let us reconsider Eq. (\ref{equality-1a}). For small $m_{1},\ m_{2}$ we
have: 
\begin{equation}
m^{2}=2\left\vert \overrightarrow{k}\right\vert \left\vert \overrightarrow{q}%
\right\vert \left( 1-\cos \varphi \right) .  \label{equality-1b}
\end{equation}%
Imposing the transverse momentum conservation as in \cite{Smith1983} or,
 for negligible $k_{3},\ q_{3}$, we get the expression for the
classical transverse mass:%
\begin{equation}
m_{\perp }^{2}=2\left\vert \overrightarrow{k}_{\perp }\right\vert \left\vert 
\overrightarrow{q_{\perp }}\right\vert \left( 1-\cos \varphi \right) ,
\label{equality-2}
\end{equation}%
and this is exactly the equation 1  proposed in Ref. \cite{Smith1983} to
determine the mass of the $W$ boson from the kinematic distribution of the decay
leptons \cite{Smith1983,CDF2022} (note that the neutrino from the $W$ boson
is not directly detectable and longitudinal momentum balance
cannot be imposed because of the geometry of the experiment \cite{CDF2022},
therefore the Eq. (\ref{equality-2}) is needed).

In conclusion, we identify the Hagen-Hurley spin one particle as the $W$
boson.

\section{Summary}
\label{summary}

We have demonstrated that the Hagen-Hurley equations describe the decay of a
spin-$1$ boson into two massive spin-$\frac{1}{2}$ fermions. The
Hagen-Hurley boson is unstable, as manifested by inequalities (\ref{INEQ}%
). Instability of the boson is enhanced by the presence in the coupled Dirac equations (%
\ref{HH2}) of non-standard higher-order spinors $\zeta _{A\dot{B}}$, $\chi _{%
\dot{C}\dot{D}}$, rather than Dirac bispinors $\left( \xi _{A},\ \eta _{\dot{%
B}}\right) ^{T}$. Moreover, in our formalism, the classical kinematic
condition for the transverse mass (\ref{equality-2}) appears naturally.

Summing up, we think that the Hagen-Hurley equations in spinor form (\ref{HH}%
) \cite{Lopuszanski1978} or in alternative $7\times 7$ matrix formulation 
\cite{Beckers1995a} can describe the extremely unstable intermediate vector $%
W$ boson. Moreover, the decay of $W^{+}$ into two massive spin -- $\frac{1}{2}$
fermions ($W^{+}\longrightarrow l^{+}\nu $, $W^{+}\longrightarrow $ hadrons)
as can be described by our formalism (with massive neutrinos), amounts to
almost $100\%$ of all decay modes of the $W$ boson \cite{Workman2022}. It is
interesting, that our approach, based on the de Broglie method of fusion 
\cite{deBroglie1943,Corson1953}, also describes the kinematics of the decay.

\end{document}